\documentclass[usenatbib,useAMS]{mn2e}
\usepackage{graphicx}
\usepackage{color}  

\def\curl{{\rm curl}\,}
\def\grad{{\rm grad}\,}
\def\div{{\rm div}\,}
\def\<{\langle}
\def\>{\rangle}

\def\bi{\bf}

\def\be{\begin{equation}}
\def\ee{\end{equation}}
\def\bea{\begin{eqnarray}}
\def\eea{\end{eqnarray}}
\def\nn{\nonumber}
\newcommand{\ms}{\noalign{\vspace{3pt plus2pt minus1pt}}}

\title{Non-corotating models for pulsar magnetospheres}
\author[D. B. Melrose and R. Yuen]
{D. B. Melrose$^1$ and R. Yuen$^{1,2}$\\
$^1$SIfA, School of Physics, University of Sydney, Sydney, NSW 2006, Australia\\
$^2$CSIRO Astronomy and Space Science, Australia Telescope National Facility, P.O. Box 76, Epping, NSW 1710, Australia}
\date{%Received in original form \hspace{2cm}}
          --- Accepted for publication}
\pubyear{213}

\begin{document}

\maketitle
\begin{abstract}
We reconsider pulsar electrodynamics for an obliquely rotating pulsar, and propose a way of synthesizing the vacuum dipole model (VDM) and the rotating magnetosphere model (RMM). We first modify the VDM by assuming that the parallel component of the inductive electric field is screened by charges. We refer to the resulting model as the minimal model. We calculate the screening charge density in the minimal model and compare it with the (Goldreich-Julian) charge density in the RMM. We identify the plasma velocity in the minimal model as the electric drift velocity due to the perpendicular component of the inductive electric field. We define a class of synthesized models as a linear combination of a fraction $y$ times the minimal model and $1-y$ times the RMM. These models require a gap (with $E_\parallel\ne0$) between the corotating stellar surface and the non-corotating magnetosphere. We present illustrative plots of the charge density, of the location of nulls (where the charge density is zero) and of the three components of the plasma velocity as a function of the angles ($\theta,\psi$) relative to the rotation axis, for specific values of the obliquity $\alpha$ and the parameter $y$. We discuss the question ``Can any pulsar magnetosphere be corotating?'' critically, pointing out difficulties associated with setting up corotation in the polar cap region. We speculate that the corotating plasma may flow across the last closed field line from the closed-field region. We suggest that abrupt changes in the spin-down rate in some pulsars may be due to jumps between the RMM and the minimal model. 

\end{abstract}


\section{Introduction}
There is now compelling evidence that abrupt changes in the magnetosphere of some pulsars can alter their slowing down rate. The initial identification of a decrease in the slowing down rate during episodes of nulling \citep{Ketal06} in one pulsar, has been found to be a more general feature of a wider class of pulsars, linking the slowing down rate to mode changing and timing noise \citep{LH,Cetal12,Letal12}, and to changes in the X-ray emission \citep{Hetal13}.  The observations imply that the slowing-down torque on the star can change abruptly due to changes in the magnetosphere that affect both the radio emission and the X-ray emission. In polar-cap models, the main source of plasma is through pair creation on the `open' field lines that define the polar cap, and the radio emission and pulsed high-energy are assumed to originate from these open field lines. The relativistic pairs escape to beyond the light cylinder distance in a time of order the rotation period, $P$. An abrupt change in the magnetosphere can be attributed to an abrupt change in the pair creation \citep{T10b}, e.g., in its efficiency or location. The pulsar evidently jumps between two states, with different rates or locations of pair creation. \cite{Ketal06} suggested that these two states are an `on' state in which pair creation is effective, leading to a rotating magnetosphere model (RMM), and an `off' state, in which pair creation is ineffective, leading to a vacuum-dipole model (VDM). This interpretation highlights a deficiency in our present understanding of pulsar electrodynamics. The RMM and VDM are idealizations. We do not know the conditions under which either model should apply, and we have no intermediate models that would indicate how one model can jump to the other. In this paper we consider how the VDM and the RMM may be modified and synthesized to provide a range of models between which a pulsar may jump when it mode-changes or has a null. 

A notable difference between the VDM and the RMM is how the electric field is determined. In the VDM \citep{D55} the pulsar is regarded as a rotating magnetized star surrounded by a vacuum, and the magnetic and electric fields are due only to currents and charges inside and on the surface of the star. In the RMM \citep{GJ69} it is assumed that the star is surrounded by plasma that is corotating with the star. The simplest form of the RMM is for an aligned model,  $\sin\alpha=0$, where $\alpha$ is the angle between the rotation and dipole axes. There is then no time dependence, and the electrodynamics reduces to electrostatics. In the subsequent development of detailed models, the aligned case was emphasized, and the electrostatic assumption was extended to an oblique rotator, $\sin\alpha\ne0$, by postulating that the magnetosphere plasma is stationary in the corotating frame \citep{SAF78}. Thus, in the VDM the electric field is determined by Faraday's equation, and in the RMM the electric field follows from Poisson's equation.

The VDM and RMM continue to be used for different purposes. The power radiated in the VDM is assumed to be balanced by the loss of rotational energy, providing an estimate of the surface magnetic field, $\propto(P{\dot P})^{1/2}$, and the characteristic age, $P/2{\dot P}$. Electrostatic versions of the RMM are the basis for detailed modeling of the plasma in the inner regions of a pulsar magnetosphere, including acceleration and pair creation in gaps, and associated screening of the electric field. The slowing down in the RMM is attributed to the angular momentum carried away in a pulsar wind, with the stress communicated back to the star by a field-aligned current that closes across field lines in the wind and inside the star \citep{S91}. 

Corotation is postulated in the RMM, but this postulate is not well justified. One reason is that a nonzero parallel electric field, $E_\parallel\ne0$, is necessary and its presence precludes strict corotation. This was already  incorporated into the \cite{RS75} model for an aligned rotator. In brief, the corotation charge density can be maintained only if there is a magnetospheric source of charge,  $E_\parallel\ne0$ is needed to accelerate  primary particles to very high energy triggering a pair cascade that provides the required source of charge, and in the acceleration region, called a (vacuum) gap, $E_\parallel\ne0$ implies that the frozen-in condition does not apply, so that plasma slips relative to the magnetic field within the gap. In the \cite{RS75} model the polar-cap region  rotates at an angular speed $\omega'$ that is less than the angular speed, $\omega_*$,  of the star, with the change occurring across the gap and related to the parallel potential drop determined by the integral of $E_\parallel$ across the gap. For the same reason, a gap with $E_\parallel\ne0$ is required in the polar cap regions of an oblique rotator, and the magnetosphere above the gap cannot be corotating with the star. A second reason applies specifically to an oblique rotator. The electric drift associated with the corotation electric field determines only the component of the corotation velocity across the magnetic field. The corotation velocity also has a component along the magnetic field lines. For example, in the case of an orthogonal rotator the parallel component is $\omega_*r\sin(\omega_*t)$. This changing parallel velocity requires a mechanical force to accelerate and decelerate plasma along field lines. In the context of planetary magnetospheres, \cite{HB65} suggested a force associated with the bounce motion of trapped particles, which reflect at moving mirror points. This force is not relevant in polar-cap models in which particles escape freely along open field lines. There appears to be no mechanical force that can accelerate the plasma along field lines in the open-field region. (An apparent counter-argument is that force-free models for pulsar magnetosphere quickly adjust to corotation \citep{S06}, but these models do not include the gap and associated magnetospheric source of plasma that are central to our argument.) The arguments against strict corotation in the open-field regions are compelling. However, it is unclear how the plasma motion is determined when the corotation assumption is abandoned.

To synthesize the VDM and the RMM one needs to include the effect of the plasma, through its charge and current densities ($\rho$ and ${\bi J}$, respectively) in the VDM.  In Section~\ref{section:response} we adopt the same procedure that we used previously  \citep{MY12}, based either on drift motions (orbit theory) or on cold-plasma response theory.  A different approach is adopted in ``force-free'' models \citep{K02,C05,S06,Li_etal12a,Li_etal12b,P12}, which are based on a fluid approach and a form of Ohm's law. Strict corotation is implied by assuming infinite conductivity, $\sigma\to\infty$, which implies $E_\parallel=0$. The condition $\sigma\to\infty$ is relaxed in the intermediate model of \cite{Li_etal12a,Li_etal12b} by allowing a region with $\sigma\to0$, and hence with $E_\parallel\ne0$. In our approach, we assume that the $E_\parallel$ implied by the VDM cannot be perfectly screened, and that the degree to which it is screened can change, described by our parameter $y$. The relation between these two approaches for including $E_\parallel\ne0$ is unclear. We comment on other differences in these two approaches in  Section~\ref{section:response}, and in more detail in  Section~\ref{section:corotate}.

In seeking to synthesize the VDM and the RMM we concentrate on the electric fields in the two models. We make the following points:
\vspace{-6pt}
\begin{enumerate} 
\item The important electric field  in the VDM is inductive ($\curl{\bi E}=-\partial{\bi B}/\partial t\ne0$, $\div{\bi E}=0$), and it includes a component, $E_\parallel$, along the magnetic field.
\item It is impossible to screen an inductive electric field by charges, but it is possible to screen $E_\parallel$ locally by charges.
\item There is a (quadrupolar) potential electric field in the VDM, due to a surface charge on the star, and ``screening of the vacuum field'' usually refers only to this field.
\item The electric field  in the RMM is the corotation field,  which is electrostatic  ($\curl{\bi E}=0$, $\div{\bi E}\ne0$) for $\alpha=0$.
\item For $\alpha\ne0$, the corotation field can be separated into the inductive field plus a potential field.
\item A rotating ${\bi B}$ does not imply corotating plasma.
\end{enumerate}
\vspace{-6pt}
We amplify on most of these points in this paper, commenting first on the last point.

As we show in section~\ref{section:Efields},  the electric and the magnetic fields in both the VDM and the RMM all obey the equation of motion for a rotating vector field. It is well known that the frozen-in condition implies that when the plasma moves with velocity ${\bi u}$, the magnetic field satisfies $\partial{\bi B}/\partial t=\curl({\bi u}\times{\bi B})$. However, the reverse is not valid. The motion of field lines is not uniquely determined by the equation of motion for ${\bi B}$ \citep{B78}, which does not imply the frozen-in condition. This point was made in the present context by \cite{M71}, who noted that the equation of motion for a magnetic field moving at the corotation velocity implies that the electric field is the corotation field plus an arbitrary potential field, which implies an arbitrary additional electric drift velocity.

A seemingly obvious way to synthesize the VDM and the RMM (with $\alpha\ne0$) is to take a linear combination of the two models, with the expectation that this would lead to a model with plasma rotating at $\omega'<\omega_*$.  However, this is not correct. Moreover, it leads to a conceptual difficulty because the inductive $E_\parallel\ne0$ implies that the frozen-in condition does not apply. The ``plasma velocity'' is not meaningfully defined for $E_\parallel\ne0$. To overcome this difficulty, we introduce a ``minimal'' model in which it is assumed that the inductive $E_\parallel$ is screened by charges. The idea behind this assumption is strongly dependent on the method we use to include the plasma response, cf.\ Section~\ref{section:response}. In contrast with force-free models, in which the parallel response is assumed to be dissipative, we note that the reactive part of the parallel response is strongly oscillatory, leading to the development of large-amplitude electric oscillations  \citep{Letal05,BT07,T10a}. We assume that, with the exception of localized gap-like regions, the oscillating $E_\parallel$ averages to zero over many oscillations. The assumption $E_\parallel=0$ implies that the plasma velocity across the field lines is well defined (except in gaps). In the minimal model, the plasma velocity corresponds to the electric drift velocity due to the perpendicular component of the inductive electric field \citep{MY12}.

We propose a class of synthesized models for an oblique rotator by taking a linear combination of the minimal model and the RMM. We characterize this class by the fraction $y$ of the minimal model in this synthesis. In this class of models the charge density and plasma velocity are linear combinations of their values in the minimal model and the RMM.  We present numerical results for the charge density, location of null (in the charge density) points, and the plasma velocity across the magnetic field for several choices of the parameters $\alpha$ and $y$.  Null points play an important role in the theory, with the argument being that the charge density can change sign along a field line only if there is a local source of charges, leading to the concept of an outer gap \citep{CHR85}. Both high-energy emission, and radio emission are possibly associated with pairs created in association with an outer gap. The plasma velocity is potentially observable through the drift of subpulses through the pulse window. 

We suggest jumps  between the RMM ($y=0$) and the minimal model ($y=1$) as a possible interpretation for the observed abrupt changes in pulsar emission properties and associated torque.

In section~\ref{section:Efields}, we write down known expressions for the electric fields in the VDM and RMM, and point out that they all satisfy the equation of motion for a rotating vector field. In section~\ref{section:response} we describe how the plasma response may be included in the VDM, and we introduce the concept of screening of $E_\parallel$. In section~\ref{section:minimal} we describe the minimal model. In section~\ref{section:synthesized} we define the synthesized model, and in section~\ref{section:illustrative} we present plots based on our numerical calculations of the model. In section~\ref{section:corotate} we discuss the question ``Can any pulsar magnetosphere be corotating?''. Our conclusions are summarized in section~\ref{section:conclusions}.

\section{Rotating fields} 
\label{section:Efields}

In this section we point out that all fields in the VDM and the RMM are rotating vector fields.

\subsection {Rotating vector field}

The equation of motion for a vector field, ${\bi V}$, rotating with angular velocity $\bomega_*$ is \citep{M67}
\bea
{\partial{\bi V}\over\partial t}=-(\bomega_*\times{\bi x})\cdot\grad{\bi V}
+\bomega_*\times{\bi V}
\qquad\qquad
\nn
\\
=\curl[(\bomega_*\times{\bi x})\times{\bi V}]
+(\bomega_*\times{\bi x})\div{\bi V},
\;\;\,\,
\label{RVF}
\eea
where the second form follows from the first by standard vector identities. All the fields in the VDM and RMM, which satisfy the Maxwell's equations
\be
\curl{\bi E}=-{\partial{\bi B}\over\partial t},
\qquad
\curl{\bi B}=\mu_0\left({\bi J}+\varepsilon_0{\partial{\bi E}\over\partial t}\right),
\label{pot3}
\ee
also satisfy the equation of motion (\ref{RVF}), as we now show.

\subsection{Time-dependent magnetic dipole}

The calculation of the magnetic and electric fields due to a time-dependent magnetic dipole, ${\bf m}(t)$, in vacuo is a textbook problem. Introducing the retarded time $t_{\rm ret}=t-r/c$, where $r=|{\bi x}|$ is the radial distance from the center of the star, the vector potential is
\be
{\bi A}(t,{\bi x})
={\mu_0\over4\pi}\left[
-{{\bi x}\times{\bi m}(t_{\rm ret})\over r^3}
-{{\bi x}\times{\dot{\bi m}(t_{\rm ret})}\over r^2c}\right],
\label{Eind1}
\ee
where a dot denotes a time derivative. The electric and magnetic fields are determined by
 \be
{\bi E}(t,{\bi x})=-{\partial{\bi A}(t,{\bi x})\over\partial t},
\qquad
{\bi B}(t,{\bi x})=\curl {\bi A}(t,{\bi x}).
\label{Eind2}
\ee
The electric field is, omitting arguments,
 \be
{\bi E}={\mu_0\over4\pi}\left[
{{\bi x}\times{\dot{\bi m}}\over r^3}
+{{\bi x}\times{\ddot{\bi m}}\over r^2c}\right].
\label{Eind3}
\ee 
The terms $\propto1/r^2$ and $1/r$ are referred to as the inductive and radiative terms, respectively. The magnetic field is
\bea
{\bi B}={\mu_0\over4\pi}
\bigg[{3{\bi x}\,{\bi x}\cdot{\bi m}-r^2{\bi m}\over r^5}+{3{\bi x}\,{\bi x}\cdot{\dot{\bi m}}-r^2{\dot{\bi m}}\over r^4c}
\nn
\\
\qquad
+{{\bi x}\times({\bi x}\times{\ddot{\bi m}})\over r^3c^2}
\bigg].
\qquad\qquad
\label{Eind4}
\eea
The terms $\propto1/r^3,1/r^2$ and $1/r$ are referred to as the dipolar, inductive and radiative terms, respectively. 

For a rotating field one has
\be
{\dot{\bi m}}=\bomega_*\times{\bi m},
\qquad
{\ddot{\bi m}}=\bomega_*\times(\bomega_*\times{\bi m}).
\label{rf3}
\ee
It is straightforward to show that ${\bi m}(t-r/c)$, and the expressions for ${\bi A}$, ${\bi E}$ and ${\bi B}$ from (\ref{Eind1})--(\ref{Eind4}) with (\ref{rf3}), all satisfy (\ref{RVF}). Hence all are rotating vector fields. For ${\bi B}$, 
\be
{\partial{\bi B}\over\partial t}=\curl[(\bomega_*\times{\bi x})\times{\bi B}],
\label{RVB}
\ee 
is satisfied separately for each of the terms $\propto1/r^3,1/r^2$ and $1/r$ in (\ref{Eind4}). These are vacuum fields, and it does not follow from (\ref{RVB}) that any plasma present would corotate. As we have already shown \citep{MY12}, the electric drift motion due to the vacuum fields is quite different from the corotation velocity $\bomega_*\times{\bi x}$.

\subsection{Corotation field}

The corotation electric field,
\be
{\bi E}_{\rm cor}=-(\bomega_*\times{\bi x})\times{\bi B},
\label{cf1}
\ee
for an obliquely rotating magnetosphere, may be written in the form \citep{HB65,M67}
\be
{\bi E}_{\rm cor}=-\grad\Phi_{\rm cor}-{\partial{\bi A}\over\partial t},
\label{cf3}
\ee
with \citep{M67}
\be
\grad\Phi_{\rm cor}=(\bomega_*\times{\bi x})\cdot{\bi A}+\Phi_0.
\label{cf4}
\ee
For ${\bi A}$ corresponding to a rotating dipole, given by (\ref{Eind1}), one needs $\Phi_0$ to cancel the Coulomb potential implied by the average over directions, $\langle(\bomega_*\times{\bi x})\cdot{\bi A}\rangle$, of the first term in (\ref{cf4}). One then finds
\be
\Phi_{\rm cor}={\mu_0\over4\pi}
\left(-{\bomega_*\cdot{\bi x}\,{\bi m}\cdot{\bi x}\over r^3}+{\bomega_*\cdot{\bi m}\over3r}
\right).
\label{cf5}
\ee

It was pointed out by \cite{HB65} that the potential field in (\ref{cf3}) screens the parallel component of the inductive field. This is obvious from the fact that ${\bi E}_{\rm cor}$ in the form (\ref{cf1}) has zero component along ${\bi B}$, and hence the parallel component of the potential and inductive terms on the right hand side of (\ref{Eind3}) must cancel. However, the potential field in the RMM is not parallel to ${\bi B}$, and hence does not just screen the parallel inductive electric field, but also modifies the perpendicular electric field.

\cite{M71} derived the corotation field by combining the equation of motion (\ref{RVB}) for a rotating magnetic field and Faraday's law, writing the solution in the form
\be
{\bi E}+(\bomega_*\times{\bi x})\times{\bi B}=-\grad\Psi,
\label{M71}
\ee
with $\Psi$ arbitrary. The solution (\ref{M71}) reproduces the corotation field (\ref{cf1}) for $\Psi=0$. For the choice \citep{E72} $\Psi=-\Phi_{\rm cor}$, (\ref{M71}) gives the vacuum electric field, ${\bi E}=-\partial{\bi A}/\partial t$, where the identity
\be
{\partial{\bi A}\over\partial t}=(\bomega_*\times{\bi x})\times{\bi B}-\grad\Phi_{\rm cor}
\label{cf7}
\ee
is used. It follows that the VDM and the RMM are limiting cases of (\ref{M71}) with $\Psi=-\Phi_{\rm cor}$ and $\Psi=0$, respectively.

\subsection{Goldreich-Julian charge density}

The divergence of ${\bi E}_{\rm cor}$ determines the corotation charge density,
\be
\rho_{\rm cor}=\varepsilon_0\div{\bi E}_{\rm cor}
=-2\varepsilon_0\bomega_*\cdot{\bi B}
+\varepsilon_0(\bomega_*\times{\bi x})\cdot\curl{\bi B}.
\label{cf2}
\ee
The Goldreich-Julian charge density \citep{GJ69}, which was derived for an aligned rotator, follows from (\ref{cf2}) by assuming $\curl{\bi B}=\mu_0\rho_{\rm cor}\bomega_*\times{\bi x}$, and solving to find 
\be
\rho_{\rm cor}=\rho_{\rm GJ}=-{2\varepsilon_0\bomega_*\cdot{\bi B}\over1-|\bomega_*\times{\bi x}|^2/c^2}.
\label{GJ1}
\ee

The result (\ref{GJ1}) also applies in the oblique case. To show this, we note that the displacement current associated with the corotation electric field (\ref{cf1}) follows from
\be
{\partial{\bi E}_{\rm cor}\over\partial t}=(\bomega_*\times{\bi x})\times\curl{\bi E}_{\rm cor}.
\label{A1}
\ee
We include the displacement current (\ref{A1}) in the second Maxwell equation (\ref{pot3}), along with the actual current ${\bi J}=\rho\,\bomega_*\times{\bi x}$, and neglect all other currents. On inserting the resulting expression for $\curl{\bi B}$ into (\ref{cf2}), the result (\ref{GJ1}) follows.

\section{Including the plasma response}
\label{section:response}

In this section we show how the plasma response can be included in an electrodynamic model.

\subsection{Electric drift velocity}

The electric field in the magnetosphere may be separated into components perpendicular and parallel to the magnetic field:
\be
{\bi E}={\bi E}_\perp+E_\parallel{\bi b},
\label{E1}
\ee
with ${\bi b}={\bi B}/B$ the unit vector along the (dipolar) field line. For the displacement current one has
\be
{\bi J}_{\rm displ}=\varepsilon_0{\partial{\bi E}\over\partial t}
=\varepsilon_0{\partial{\bi E}_\perp\over\partial t}
+\varepsilon_0{\partial(E_\parallel{\bi b})\over\partial t}.
\label{E2}
\ee

Provided that ${\bi E}_\perp$ is changing slowly in time (the characteristic frequency associated with the change is small in comparison with the plasma and cyclotron frequencies in the plasma), its effect is to cause all particles to drift across the field lines at the electron drift velocity, ${\bi v}_E$. Then (\ref{E1}) becomes
\be
{\bi E}=-{\bi v}_E\times{\bi B}+E_\parallel{\bi b},
\qquad
{\bi v}_E={{\bi E}_\perp\times{\bi B}\over B^2}.
\label{E3}
\ee
Because all particles have the same electric drift velocity, ${\bi v}_E$ may be interpreted as perpendicular component of the fluid velocity, ${\bi u}$, of the plasma. As noted by \cite{HB65}, if ${\bi u}$ is identified as the corotation velocity, $\bomega_*\times{\bi x}$, then it has a component along the field lines that cannot be explained in terms of a drift motion.

\subsection{Polarization current}

A temporally changing (perpendicular) electric field causes charges of opposite sign to drift in the direction of $\partial{\bi E}_\perp/\partial t$ at the polarization drift velocity, which is proportional to their mass to charge ratio. After summing over the contributions of all species of charged particle, this implies that the plasma responds through a polarization current density
\be
{\bi J}_{\rm pol}={c^2\over v_A^2}\varepsilon_0{\partial{\bi E}_\perp\over\partial t}.
\label{pd1}
\ee
The sum of the polarization and displacement currents appears in the second of the Maxwell equations (\ref{pot3}) in the form
\be
\mu_0{\bi J}_{\rm pol}+{1\over c^2}{\partial{\bi E}_\perp\over\partial t}={1\over v_0^2}{\partial{\bi E}_\perp\over\partial t},
\qquad
v_0^2={v_A^2\over1+v_A^2/c^2},
\label{pd2}
\ee
where $v_0$ is the MHD speed. In a pulsar magnetosphere one has $v_A^2\gg c^2$ and the MHD speed is close to the speed of light, $v_0\approx c$. It follows that the inclusion of the plasma response to $\partial{\bi E}_\perp/\partial t$ leads to a correction of order $c^2/v_A^2\ll1$.

Neither the electric drift nor the polarization current require that the charges have a nonzero pitch angle, and both apply to electrons and positrons in their ground state in a pulsar magnetosphere.

\subsection{Cold plasma response}

Although a cold plasma is a poor approximation for the (one-dimensional, relativistic pair) plasma in a pulsar magnetosphere, it suffices to identify the form of the response under more general conditions. 

The response of the plasma may be described by the induced (ind) charge and current densities, which may be included in the polarization, ${\bi P}$, and the electric induction:
\be
{\bi D}=\varepsilon_0{\bi E}+{\bi P},
\qquad
\rho_{\rm ind}=-\div{\bi P},
\qquad
{\bi J}_{\rm ind}={\partial{\bi P}\over\partial t}.
\label{pd5}
\ee
We are concerned with the response of the plasma at frequencies much lower than the natural frequencies, which are the plasma frequency, $\omega_p$, and the cyclotron frequencies of each species of particle. A derivation of the low frequency limit of the cold-plasma response is outlined in Appendix~\ref{appendix:cold}. The response to the perpendicular and parallel components of the electric field are different.

The perpendicular component of the response includes a Pedersen-like term
\be
{\bi P}_\perp={c^2\over v_A^2}\varepsilon_0{\bi E}_\perp.
\label{pd6_1}
\ee
This implies an induced current
\be
{\bi J}_{\rm ind \perp}={c^2\over v_A^2}\varepsilon_0{\partial{\bi E}_\perp\over\partial t},
\label{pd7_1}
\ee
which is the polarization current (\ref{pd1}). There is also a Hall-like term
\be
{\bi J}_{\rm ind\,H}=\rho_0{{\bi E}\times{\bi B}\over B^2},
\label{pd9}
\ee
which, in the RMM, is the space-charge current included in the derivation of (\ref{GJ1}). The parallel response in the low-frequency, cold-plasma limit can be derived from Newton's equation $mdv_\parallel/dt=qE_\parallel$ for a charge $q$ and mass $m$ by multiplying by $qn/m$, where $n$ is the number density, and summing over all species of particles. This gives
\be
{\partial J_{\rm ind\parallel}\over\partial t}=\varepsilon_0\omega_p^2E_\parallel-\nu_{\rm eff} J_{\rm ind\parallel},
\label{pd8_1}
\ee
where a dissipative term is included through an effective collision frequency, $\nu_{\rm eff} $.

The parallel response (\ref{pd8_1}) in Maxwell's equations implies that $E_\parallel$ satisfies
\be
\left({1\over c^2}{\partial^2\over\partial t^2}+{\omega_p^2\over c^2}-\nabla_\perp^2\right)E_\parallel=
\mu_0\nu_{\rm eff}J_\parallel-{\partial\over\partial z}{\rm div}_\perp{\bi E}_\perp,
\label{depar}
\ee
where $\mu_0\nu_{\rm eff}$ is an effective resistivity. The response is strongly oscillatory, tending to lead to oscillations at $\omega_p$. Acceleration of electrons by $E_\parallel$ sets up a current, and when the current density exceeds a relevant threshold, and instability develops such that longitudinal waves grow and provide an anomalous resistivity. The low-frequency response, averaged over times $\gg1/\omega_p$ such that the term involving $\partial^2/\partial t^2$ can be neglected then becomes quasi-static, and dependent on the effective resistivity. Although the details are different, we argue that analogous processes occur in a pulsar magnetosphere, resulting in screening of $E_\parallel$.

\subsection{Oscillations in $E_\parallel$}
\label{sect:screen}

In the electron-positron plasma in a pulsar magnetosphere, the presence of $E_\parallel\ne0$ leads to growth of electrostatic oscillations to large-amplitude \citep{Letal05,BT07,T10a}. The frequency of these oscillations is at a counterpart of the plasma frequency that depends on the maximum Lorentz factor to which the electrons and positrons are accelerated in the wave, and this maximum Lorentz factor is limited by the threshold for effective pair creation. The pair production provides both an intrinsically magnetospheric source of charge, and also a dissipative effect that can plausibly be simulated by $\nu_{\rm eff}$. After averaging over the oscillations, the low-frequency response can plausibly be described by an equation of the form (\ref{depar}) with the term involving $\partial^2/\partial t^2$ omitted. We speculate that this can lead to a quasi-static screening of $E_\parallel$. To be more specific, we assume that $E_\parallel$ is screened except in localized regions where $E_\parallel\ne0$ is needed to provide additional charges, for example to allow the screening charge density to change sign. We should emphasize that this speculation has been accepted in the literature on pulsar electrodynamics since the 1970s, through the parallel screening implicit in the RMM. Nevertheless, a better justification than we are able to provide is clearly needed.

\section{Minimal model}
\label{section:minimal}

In this section we introduce a minimal model for screening in a pulsar plasma, and then define a range of intermediate models.

\subsection{Screening of $E_\parallel$}

We define a minimal model in which ${\bi E}_{\rm ind\parallel}$ is screened, and ${\bi E}_{\rm ind\perp}$ has the same value as in the VDM. This requires that there be a charge density that produces an electric field ${\bi E}_{\rm min}=-{\bi E}_{\rm ind\parallel}$, where `min' refers to the minimal model. We further assume that on the timescale ($P$) on which ${\bi E}_{\rm ind\parallel}$ changes, ${\bi E}_{\rm min}$ also changes, so that the displacement current is also screened, $\partial{\bi E}_{\rm min}/\partial t=-\partial{\bi E}_{\rm ind\parallel}/\partial t$. 

The generalization from the VDM to the minimal model does not affect the electric drift velocity, ${\bi v}_{\rm ind}={\bi E}_{\rm ind\perp}\times{\bi B}/B^2$, associated with ${\bi E}_{\rm ind\perp}$, but it does affect its interpretation. The frozen-in condition is not satisfied in the VDM, and the drift velocity can be interpreted only as the drift velocity of any test charge. With $E_\parallel=0$ in the minimal model, motion of field lines becomes meaningful, and the field lines move at ${\bi v}_{\rm ind}$. Note the counter-intuitive implication: in the minimal model the magnetic field satisfies the equation of motion for a rotating magnetic field, but the motion of the field lines, frozen-in to the plasma, is quite different from the corotation velocity \citep{MY12}.

The divergence of the screening electric field implies a screening charge density
\be
\rho_{\rm min}=-\varepsilon_0\div({\bi b}E_{\rm ind\parallel}).
\label{sc1}
\ee
The calculation of $\rho_{\rm min}$ is complicated by the fact that the field lines are curved. The details are outlined in appendix~\ref{appendix:min}. The result is
\bea
\rho_{\rm min}=-{m\omega_*\over4\pi r^3c^2}\, \frac{18 \cos\theta_m(1+\cos^2\theta_m)}{(1 + 3 \cos^2\theta_m)^2}
\qquad\qquad
\nn
\\
\times
 (\cos\theta - \cos\theta_m \cos\alpha),
\label{sc10}
\eea
where we introduce the polar angle $\theta_m$ between the position vector, ${\bi x}$, and the dipole axis, and write
\be
\cos\theta_m=\cos\alpha\cos\theta+\sin\alpha\sin\theta\cos(\phi-\omega_* t),
\label{thetam}
\ee
where ${\bi m}$ is assumed to be in the plane $\phi=0$ at $t=0$. The angles are defined by writing
\be
{{\bi x}\cdot{\bi m}\over rm}=\cos\theta_m,
\quad
{{\bi x}\cdot\bomega_*\over r\omega_*}=\cos\theta,
\quad
{{\bi m}\cdot\bomega_*\over m\omega_*}=\cos\alpha,
\label{sc9}
\ee
For comparison, the Goldreich-Julian value, without the denominator in (\ref{GJ1}), is
\be
\rho_{\rm GJ}=-{m\omega_*\over 2\pi r^3c^2}\,(3\cos\theta\cos\theta_m-\cos\alpha).
\label{sc11}
\ee

\subsection{Location of nulls}

A qualitative difference between the minimal model and the RMM concerns the locations of the null points, where the charge density is zero.  The original argument for the outer gap \citep{CHR85} is that pair creation must be important near a null to allow the charge density to change sign along a field line. In an aligned model one has $\cos\alpha=1$, $\theta_m=\theta$, and (\ref{sc11}) implies that the null surface in the RMM is at $3\cos^2\theta=1$. In the oblique form of the RMM, the null surface is at the solutions of $3\cos\theta\cos\theta_m=\cos\alpha$. In the minimal model, the charge density (\ref{sc10}) implies nulls at the solutions of $\cos\theta=\cos\alpha\cos\theta_m$ and of $\cos\theta_m=0$.

\subsection{Screening current density}

An intrinsically different feature of an oblique model, compared with an aligned model, is that there is a screening current density. This may be seen by noting that the screening charge density changes as a function of time, and the equation of charge continuity, $\partial\rho/\partial t+\div{\bi J}=0$, requires an associated ${\bi J}$. This ${\bi J}$ has not been included explicitly in any existing model.

In the minimal model it is straightforward to see that
\be
{\bi J}_{\rm min}=\varepsilon_0{\bi b}\partial E_\parallel/\partial t
\label{Jsc}
\ee 
is a solution of the charge-continuity equation. In the minimal model, by hypothesis, the parallel displacement current in the absence of screening is canceled by the equal and opposite $\varepsilon_0\partial{\bi E}_{\rm min}/\partial t$, so that there is no parallel displacement current on the right hand side of the second of the Maxwell equations (\ref{pot3}). The current (\ref{Jsc}) provides a contribution that is equal to the original displacement current. Thus, the inclusion of parallel screening does not change the value of the right hand side of the second of the Maxwell equations (\ref{pot3}). The  parallel displacement current in the absence of screening is replaced by an equal contribution from the current (\ref{Jsc}) carried by the particles in the presence of screening.

An analogous result applies to the (oblique) RMM. The proof generalizes to any electric field written in terms of potentials, $\Phi$ and ${\bi A}$, in the Coulomb gauge, $\div{\bi A}=0$,
\be
{\bi E}={\bi E}_{\rm pot}+{\bi E}_{\rm ind},
\quad
{\bi E}_{\rm pot}=-\grad\Phi,
\quad
{\bi E}_{\rm ind}=-{\partial{\bi A}\over\partial t}.
\label{pot1}
\ee
The divergence of ${\bi E}$ implies the charge density, $\rho_{\rm pot}$, and the solution
\be
\rho_{\rm pot}=\varepsilon_0\,\div{\bi E}_{\rm pot},
\qquad
{\bi J}_{\rm pot}=-\varepsilon_0{\partial{\bi E}_{\rm pot}\over\partial t},
\label{pot2}
\ee
is consistent with the charge-continuity equation. In the second of the Maxwell equations (\ref{pot3}), the combination ${\bi J}_{\rm pot}+\varepsilon_0\partial{\bi E}_{\rm pot}/\partial t$ is identically zero. As in the minimal model, the right hand side of the second of the Maxwell equations (\ref{pot3}) is the same as in the absence of screening, but with the displacement current replaced by an identical screening current.

One implication is that inclusion of the screening does not alter the solution of the Maxwell equations from the inductive field {\it in vacuo}. Specifically, for a rotating dipole, equations (\ref{Eind1})--(\ref{Eind4}) correctly include the time-dependence of the inductive field even in the presence of minimal screening. This result applies only to the effect of the screening current. Other currents must be balanced by $\curl{\bi B}$ in (\ref{pot3}). In particular, in force-free models the current density modifies the background magnetic field substantially through $\curl{\bi B}\ne0$ near and beyond the light cylinder. The new point we are making is that the current density associated with the time-varying charge density does not affect the background magnetic field.

\section{Synthesized model}
\label{section:synthesized}

We define a synthesized model as a linear combination of the minimal model and the RMM, specifically, $y$ times the minimal model plus $(1-y)$ times the RMM. This model has an electric field
\be
{\bi E}=(1-y{\bi b}\,{\bi b}\cdot){\bi E}_{\rm ind}
-(1-y)\grad\Phi_{\rm cor},
\label{syn2}
\ee
a charge density
\begin{equation} \label{eq:SynModelChargeDensity}
\rho_{\rm sn} = y\, \rho_{\rm min} + (1-y)\, \rho_{\rm GJ}, 
\end{equation}
and a drift velocity 
\be
{\bi v}_{\rm dr}=y\,{\bi v}_{\rm ind}
+(1-y)\,(\bomega_*\times{\bi x})_\perp.
\label{drift}
\ee
For $y=1$ the model reduces to the minimal model. For $y=0$ the model corresponds to the RMM, except that, as already remarked, some additional physics is needed to provide the parallel component of the corotation velocity $\bomega_*\times{\bi x}$ \citep{HB65}.

%We suggest that a plausible model for the two states between which a pulsar magnetosphere may jump abruptly correspond to $y=0$ and $y=1$ in (\ref{syn2})--(\ref{drift}).

\begin{figure}
\begin{center} 
\includegraphics[width=.45\textwidth]{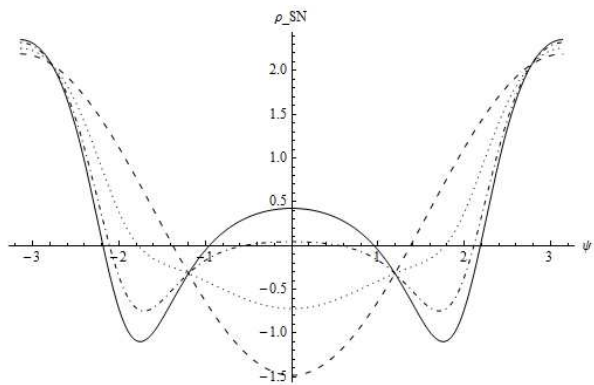}
\caption{Variations in charge density against rotation phase for $y$ = 0 (dashed), 0.4 (dotted), 0.8 (dot-dashed), and 1 (solid) at $\alpha = \pi/4$ and $\theta = \pi/3$.  The vertical axis is the charge density in units of $m\omega_* /4\pi  r^3 c ^2$. }
\label{fig-CH_A45T60Ys}
\end{center}
\end{figure}

\begin{figure}
\begin{center} 
\includegraphics[width=.45\textwidth]{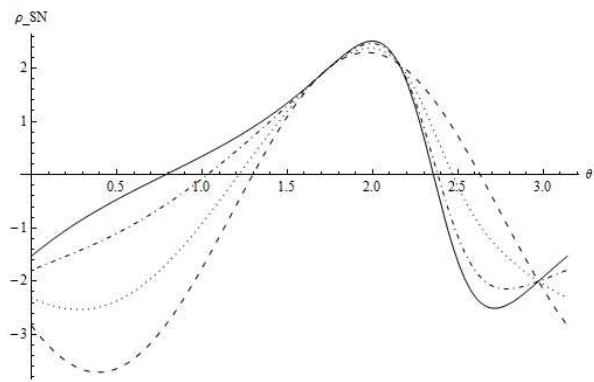}
\caption{The screening charge density for $\alpha= \pi/4$ and $\psi = 0$ is plotted as a function of $0<\theta<\pi$ for the minimal model (solid), the RMM (dashed) and two intermediate states correspond to $y = 0.4$ (dotted) and $y=0.8$ (dot-dashed).}
\label{fig-ChargeDensitiesForYs}
\end{center}
\end{figure}

\section{Illustrative examples}
\label{section:illustrative}

In this section we present plots illustrating the charge density and drift velocity for specific examples of the synthesized model in (\ref{syn2})--(\ref{drift}). These plots include the RMM and the minimal model as the special cases $y=0$ and $y=1$, respectively. We are not concerned with the absolute value of the charge density, or of the electric fields, and in this section we set $m = \omega_* = r = c = 1$.

\subsection{Charge density}

We evaluate and plot the charge density (\ref{eq:SynModelChargeDensity}) for a dipole magnetic field, ignoring the denominator in (\ref{GJ1}). The charge density is then a function of $\alpha,\theta,\psi,y$ times $1/r^3$, with $\psi=\phi-\omega_*t$. 

In Figure~\ref{fig-CH_A45T60Ys} we plot the charge density for a specific choice of $\alpha$ and $\theta$ as a function of $-\pi<\psi<\pi$ for four choices of $y$. In Figure~\ref{fig-ChargeDensitiesForYs} we fix $\alpha$ and $\psi$ and plot the charge density as a function of $0<\theta<\pi$. These figures show that the four curves intersect at specific points. These points correspond to parameters for which the charge densities $\rho_{\rm min}$ and $\rho_{\rm GJ}$ are equal. 

\begin{figure}
\begin{center} 
\includegraphics[width=.45\textwidth]{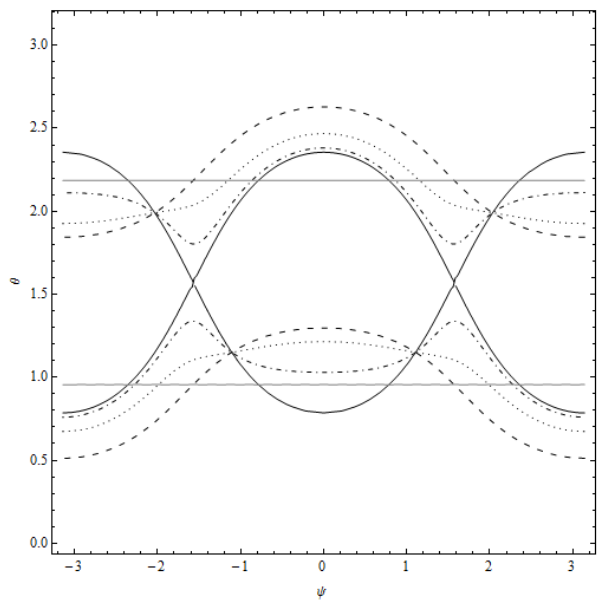}
\caption{Curves showing the location of nulls in the charge density for $\alpha= \pi/4$ as a function of $\theta,\psi$ for  the minimal model ($y = 1$, solid black), the RMM ($y = 0$, dashed) and two states correspond to $y = 0.4$ (dotted) and $y=0.8$ (dot-dashed). The two straight lines represent the nulls points for the aligned case, $\alpha=0$.}
\label{fig-NullLinesA45}
\end{center}
\end{figure}

Of particular physical interest are the points where the charge density is zero. 
The location of these null points is different for the minimal model and for the RMM, and hence depends on $y$ for the synthesized model. The location of the nulls also depends on $\alpha$ and $\psi$.  To illustrate the location of the nulls we fix $\alpha,y$ and plot the null as a curve in $\theta$--$\psi$ space. For an aligned rotator, $\alpha=0,y=0$, the nulls occur at $\cos^2\theta=1/3$ independent of $\psi$, so that the curves reduce to two lines at $\theta=\arccos1/\sqrt{3}$.  In Figure~\ref{fig-NullLinesA45} we plot the solutions for the nulls for the particular case $\alpha=\pi/4$, for the minimal model, the RMM and two intermediate cases. For smaller values of $\alpha$, the curves for the RMM and the intermediate models approach the two horizontal lines that correspond to the solution for an aligned rotator; the ratio $\rho_{\rm min}/\rho_{\rm GJ}$ goes to zero in the limit $\alpha\to0$, and the limiting case of the (solid) curve is not physically relevant. 

\begin{figure}
\begin{center} 
\includegraphics[width=.45\textwidth]{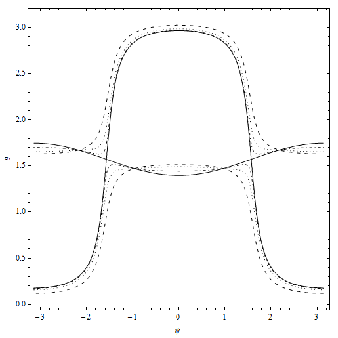}
\caption{Same as in Figure \ref{fig-NullLinesA45}, but with $\alpha = 80^\circ$.}
\label{fig-NullLinesA80}
\end{center}
\end{figure}

For larger values of $\alpha$, the curves in Figure~\ref{fig-NullLinesA45} are modified, as shown in  Figure~\ref{fig-NullLinesA80} for the case $\alpha=80^\circ$. As $\alpha$ is further increased towards the limit of an orthogonal rotator, $\alpha\to90^\circ$, the curves in Figure~\ref{fig-NullLinesA80} approach the limiting case of a horizontal line at $\theta=\pi/2$ and two vertical lines at $\psi=\pm\pi/2$. In this limit, the charge densities (\ref{sc10}) and (\ref{sc11}) both become proportional to $\cos\theta\cos\theta_m=\cos\theta\sin\theta\cos\psi$, and the lines correspond to the solutions of $\cos\theta=0$ and $\cos\psi=0$, respectively.

\subsection{Drift velocity}

\begin{figure}
\begin{center} 
\includegraphics[width=.45\textwidth]{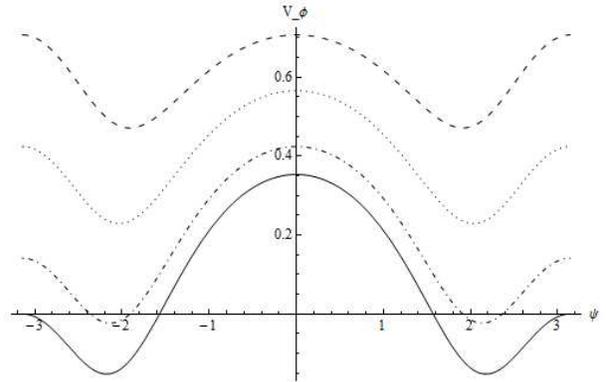}
\caption{Plot of $v_{{\rm dr},\phi}$, in units of $\omega_*r$, for $\alpha = \theta = \pi/4$ against rotation phase. The $y$ values are the same as in Figure \ref{fig-NullLinesA45}. }
\label{fig-DVp_Thetas}
\end{center}
\end{figure}

\begin{figure}
\begin{center} 
\includegraphics[width=.45\textwidth]{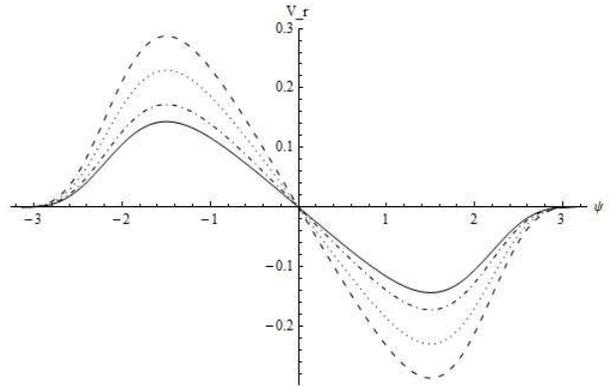}
\caption{Same as in Figure \ref{fig-DVp_Thetas}, but for the radial component $v_{{\rm dr},r}$. There is no radial component at $\psi=0$.}
\label{fig-DVr_Thetas}
\end{center}
\end{figure}

\begin{figure}
\begin{center} 
\includegraphics[width=.45\textwidth]{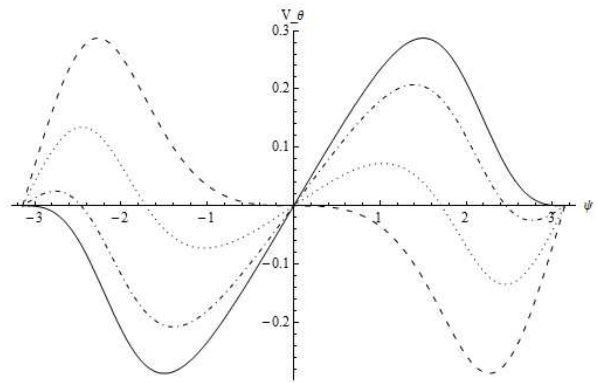}
\caption{Same as in Figure \ref{fig-DVp_Thetas}, but for the polar component $v_{{\rm dr},\theta}$.}
\label{fig-DVt_Thetas}
\end{center}
\end{figure}

The drift velocity, ${\bf v}_{\rm dr} = {\bf E} \times {\bf B}/B^2$, is  the fluid velocity of the plasma. In general it has nonzero radial, polar and azimuthal components. In Figures \ref{fig-DVp_Thetas}--\ref{fig-DVt_Thetas} we plot $v_{\rm dr,\phi}$, $v_{{\rm dr},r}$, $v_{\rm dr,\theta}$, respectively, as a function of $\psi$ for a specific value of $\alpha=\theta$ and for several choices of $y$. We choose $\alpha=\theta$ because the point $\alpha=\theta$, $\psi=0$ corresponds to the magnetic pole, at the center of the polar-cap region. The location of the point in the magnetosphere from which a distant observer sees radiation, which is assumed to be directed along field lines, corresponds to the field line that is pointing directly towards the observer.  An observer can see radiation only from a restricted range of $|\theta-\alpha|$ and $|\psi|$ about this point, with this range broadening with increasing $r$. The plots in Figures~\ref{fig-DVp_Thetas}--\ref{fig-DVt_Thetas} show that the azimuthal component of the drift velocity has an extremum at  $\alpha=\theta$, $\psi=0$, and that the radial and polar components are zero at  $\alpha=\theta$, $\psi=0$.

\begin{figure}
\begin{center} 
\includegraphics[width=.45\textwidth]{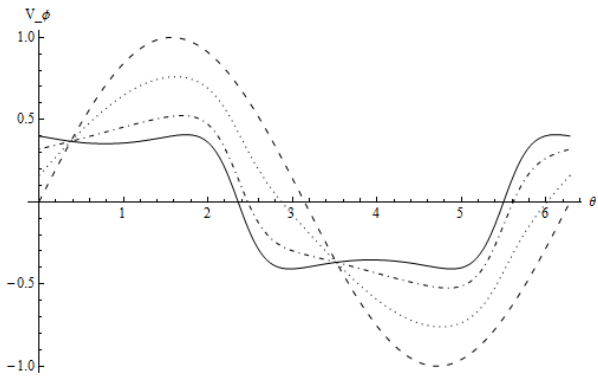}
\caption{Plot showing variations of $v_{\rm dr,\phi}$ with $\theta$ for $\alpha = \pi/4$ and $\psi = 0$. }
\label{fig-DVp_Ys_A45P0}
\end{center}
\end{figure}

The azimuthal component is of most interest for comparison with observations. It determines the apparent velocity of the plasma as it moves through the pulse window. Our choice $\theta=\alpha$ in Figure~\ref{fig-DVp_Thetas} shows how the drift velocity varies with $\psi$ within the pulse window. In Figure~\ref{fig-DVp_Ys_A45P0} we plot the azimuthal drift velocity as a function of $\theta$ for $\psi=0$. (We show values of $\theta$ from $0$ to $2\pi$, with only the range $0<\theta<\pi$ corresponding to $\psi=0$; the range $\pi<\theta<2\pi$ corresponds to $\psi=\pi$.) The plot of $v_{\rm dr,\phi}$ versus $\theta$ in Figure~\ref{fig-DVp_Ys_A45P0} shows the four curves intersecting at two points, which correspond to parameters for which $v_{\rm dr,\phi}$ has the same values for both the corotating and minimal models. 

\begin{figure}
\begin{center} 
\includegraphics[width=.47\textwidth]{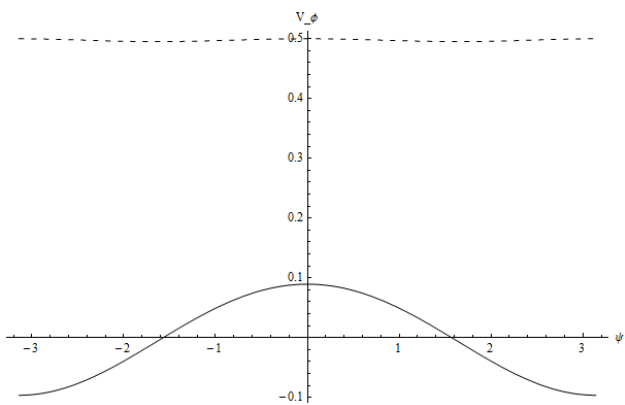}
\caption{This plot shows the variations in $v_{{\rm d},\phi}$ against $\psi$ for $\alpha = \pi/18$ and $\theta = \pi/6$ for $y = 0$ (dashing) and $y = 1$ (solid).}
\label{fig-DVp_2Ys_A10T30}
\end{center}
\end{figure}

\begin{figure}
\begin{center} 
\includegraphics[width=.48\textwidth]{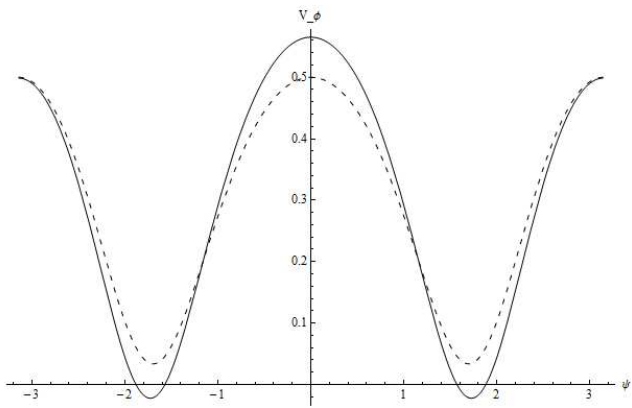}
\caption{Same as in Figure \ref{fig-DVp_2Ys_A10T30}, but for $\alpha = 4 \pi/9$.}
\label{fig-DVp_2Ys_A80T30}
\end{center}
\end{figure}

Figures~\ref{fig-DVp_2Ys_A10T30} and \ref{fig-DVp_2Ys_A80T30} show $v_{\rm dr,\phi}$ as a function of $\psi$ for $\alpha = \pi/18$ and $4 \pi/9$, respectively, and for the two choices of $y$ corresponding to the RMM and the minimal model. For the small value of $\alpha$ is chosen in Figure~\ref{fig-DVp_2Ys_A10T30}, the drift speed in the minimal model is much smaller than that in the RMM, which is nearly independent of $\psi$ and approximately equal to the corotation speed ($\omega_*r\sin\theta$). A pulsar with small $\alpha$ that jumps between the RMM and the minimal model would exhibit a large change in the drift velocity. A large value of $\alpha$ is chosen in Figure~\ref{fig-DVp_2Ys_A80T30}. The drift velocity in the minimal model is then slightly larger than the corotation velocity for $\psi\approx0$, and the velocities in both models vary approximately sinusoidal in a similar manner with $\psi$. A jump between the RMM and the minimal model would have little effect on the drift velocity in this case.

\section{Discussion}
\label{section:corotate}

It is widely assumed that a pulsar magnetosphere corotates provided there is an adequate supply of charges. In this section we discuss this assumption critically.

\subsection{Does a pulsar magnetosphere corotate?}

Corotation is a plausible assumption for most planetary and stellar magnetospheres. There are three properties of such a magnetosphere that contribute to this plausibility.\\ 
(a)  The plasma originates from the planet or star and is corotating at the surface of the planet or star.\\
(b) There is an adequate supply of charges to provide the corotation charge density and hence the potential component of the corotation electric field. \\
(c) Any deviation from corotation of trapped plasma tends to be smoothed out over many rotations.\\
Two additional requirements, specific to the oblique case, are also plausibly satisfied.\\
(d) Only the perpendicular component of the corotation velocity is associated with the corotation electric field; the parallel component of the corotation velocity requires a mechanical driver, and a form of Fermi acceleration for trapped particles has been proposed \citep{HB65}.\\
(e) The time-varying charge density implies a screening current density that has a component across the magnetic field; this current can plausibly be attributed to a polarization current driven by a much smaller (by $v_A^2/c^2\ll1$) displacement current.

None of (a)--(e) apply to the polar-cap zones of an obliquely rotating pulsar magnetosphere. The main source of plasma is pair creation well above the stellar surface \citep{T10b}, and there is no reason to expect the pairs to be corotating when they are created. This pair plasma escapes to beyond the light cylinder in a time of order a rotation period, and needs to be replaced on this time scale. There is no bounce motion to satisfy (d), and with $v_A^2/c^2\gg1$ the polarization current is ineffective in satisfying (e).

Despite these arguments, we are reluctant to reject the RMM, as a plausible first approximation, and adopt an entirely different model \citep{M04}. It is relevant to note that various arguments against corotation were discussed in connection with Jupiter's magnetosphere, but in situ data from spacecraft showed corotation to be a valid first approximation \citep{KC75,KC77}. This suggests that one needs to consider all possible mechanisms that might allow corotation to apply, at least as a first approximation, to plasma in the polar-cap regions. We discuss one possible such mechanism at the end of this section.

\subsection{Cross-field screening current}

How is the cross-field component of the screening current, required by the RMM, set up in a 1D plasma? The screening current density is written down in Appendix~\ref{appendix:Js};   the main points relevant to the discussion here are that it includes a component across field lines, and that it is required to screen the displacement current. In a plasma with $v_A^2/c^2\ll1$ a perturbation in the displacement current drives a much larger polarization current, allowing the displacement current to be screened to lowest order in an expansion in $v_A^2/c^2\ll1$. This is not possible in a pulsar magnetosphere with $v_A^2/c^2\gg1$; the polarization current is much smaller than the displacement current and cannot screen it. 

There is another way in which a screening current can effectively be driven across field lines, to maintain the required screening charge density as a function of time. Charge can flow along field lines to a conducting boundary, flow across field lines there, and flow back along different field lines. The importance of this mechanism  was recognized in a laboratory plasma by \cite{S55}; in this case, the conducting boundary is the plasma wall. The same mechanism is well accepted in the interpretation of terrestrial magnetospheric substorms \citep{McP78}, where the conducting boundary is the ionosphere, and where the closure current is referred to as a current wedge. We suggest that a similar mechanism may apply in a pulsar magnetosphere. 

An implication is that the screening current, required for corotation, can be maintained near the surface of the star, but is limited by propagation times far from the star. The screening current changes at a rate $\omega_*$ due to rotation, and a time lag $2r/c$ is involved for the propagation to and from the star to a point at radius $r$  in the outer magnetosphere.  This mechanism can maintain the screening charge density at its instantaneous value only for $2\omega_*r/c=2r/r_L\ll1$.

\subsection{Possible source of corotating plasma}

The foregoing discussion emphasizes the problems that arise with the corotation assumption for the pair plasma generated in the polar-cap region of a pulsar. In contrast, there are no strong arguments against corotation for the closed field region of a pulsar magnetosphere. Plasma flow across the last closed field line can be a source of (nearly) corotating plasma inside the polar-cap region. The inductively induced drift in (\ref{drift}) has components that allow the plasma to cross the last closed field lines. If the pulsar emission originates from near the last closed field line, as assumed in a slot-gap model \citep{H07,H08}, it is possible that the closed magnetosphere is the source of this plasma and that it is nearly corotating ($y\ll1$). Such a model needs to be investigated in detail, but we do not do so here.

\subsection{Force-free models}

The procedure adopted in Section~\ref{section:response} for including the effect of the plasma is qualitatively different from the procedure used in force-free models \citep{K02,C05,S06}, which are based on assuming $\rho{\bi E}+{\bi J}\times{\bi B}=0$. We do not attempt to compare these two different procedures in detail here, but several comments are appropriate. First, the force-free model is based on a fluid (MHD) approach in the limit in which the inertia and pressure of the plasma are negligible. Negligible inertia corresponds to the limit $c^2/v_A^2\to0$, and the differences between the two models for ${\bi J}_\perp$ involve terms that are of order $c^2/v_A^2$ and are unlikely to be important. Second, the two models involve the same ${\bi E}_\perp$, but the justifications are different. Here we assume that ${\bi E}_\perp$ is determined by Faraday's equation, and that it implies a fluid-like electric drift motion. In contrast, in the force-free approach, ${\bi E}_\perp$ is determined by Ohm's law with $\sigma\to\infty$. (This justification is not correct---the  conductivity in actually highly anisotropic, with $\sigma_\parallel\to\infty$ and $\sigma_\perp\to0$---but the conclusion on the form of ${\bi E}_\perp$ is correct.) Third, the two models  lead to quite different results for $E_\parallel$, with $E_\parallel=J_\parallel/\sigma_\parallel$ in the fluid approach, and with the relation between $E_\parallel$ and $J_\parallel$ strongly oscillatory in the approach we adopt. More generally, how one includes $E_\parallel\ne0$ correctly in any astrophysical theory remains a major unsolved problem \citep{SL06}.

A major advantage of the force-free approach is that it allows one to construct a global model, and to explore how a steady-state magnetosphere evolves from some initial condition. Of particular interest here is how the inclusion of plasma affects the slowing down rate. The slowing-down rate in a force-free model was estimated by \cite{S06} by calculating the Poynting vector (plasma inertia is neglected in force-free models).  How a change in the plasma can change the slowing-down rate requires additional assumptions. For example, \cite{T07} argued that a change in the slowing-down rate is determined by a change in the total current, \cite{BN07} argued that the slowing down is determined by the current closure on the surface of the star, and \cite{Li_etal12a} attributed the change in slowing-down to a change in the conductivity. In the synthesized model proposed here, a change in the rotational state, described by the parameter $y$, must change the slowing-down rate. However, a semiquantitative treatment, along the lines suggested by  \cite{BN07}, is beyond the scope of this paper.

\section{Conclusions}
\label{section:conclusions}

One of our objectives in this paper is to identify models for an obliquely rotating pulsar magnetosphere that are intermediate between the VDM and the RMM, and that allow jumps between different rotational states. We propose a class of synthesized models in which the rotational state is described by a parameter $0\le y\le1$. The states $y=1$ and $y=0$ are modified forms of the VDM and the RMM, respectively. A modification to the VDM is necessary for the concept of a fluid-like plasma velocity to be meaningful, which it is not in the presence of $E_\parallel\ne0$. We propose a minimal model in which the average over large-amplitude oscillations that are an inevitable consequence of $E_\parallel\ne0$  \citep{Letal05,BT07,T10a} is zero, except in localized regions that we refer to as gaps. In the regions in which the average $E_\parallel$ is zero, the concept of a fluid velocity is well-defined and equal to the electric drift velocity due to the inductive ${\bi E}_\perp$. The RMM needs to be modified because the corotation velocity, $\bomega_*\times{\bi x}$, has a (time-varying) component along the magnetic field that requires acceleration by a (non-electric) mechanical force, but no such force is available. In our synthesized model, $y=1$ corresponds to the minimal model and $y=0$ to a fluid velocity $(\bomega_*\times{\bi x})_\perp$. These models differ from each other in the form of the electric field, which can be written  as the sum either of the corotation field and a potential field \citep{M71} or of the inductive electric field and a (different) potential field \citep{M67}. The limit $y=0$ corresponds to the former of these potential fields being zero, and the limit $y=1$ corresponding to the latter potential field being that required to screen $E_\parallel$. For $0<y<1$ these models require a gap, with $E_\parallel\ne0$, as in the \cite{RS75} model for $\alpha=0$, in order to separate the corotating plasma at the surface of the star from the non-corotating plasma in the magnetosphere. Such a gap is required in any polar-cap model in which the plasma is attributed (at least in part) to pair creation in the magnetosphere. In force-free models, the need for a gap is by-passed by postulating a nonzero conductivity in the magnetosphere \citep{Li_etal12b}, which leads to a model in which $E_\parallel$ is non-zero throughout the postulated resistive region.

We suggest that the synthesized model provides a basis for understanding the abrupt  changes associated with nulling and mode changing in some pulsars. A pulsar can jump back and forth, between states corresponding to $y\approx1$ and $y\approx0$,  when it is near the threshold for effective pair creation, so that small changes can lead to the pair cascade turning on and off. A semiquantitative treatment of such changes requires further development of the model to identify the current flowing along open field lines, and how this current changes when $y$ changes.

We point out and discuss difficulties with the assumption of corotation in the oblique case. These difficulties seem overwhelming for the plasma in the polar-cap regions. Nevertheless, we are reluctant to abandon the assumption of corotation. We suggest a possible source of corotating plasma is flow across the last closed field line from the closed field region into a slot gap region. 

The main conclusion of this paper is that it is possible for the plasma in a pulsar magnetosphere to exist in different states, involving different plasma velocities, and it is possible for the pulsar to jump from one state to another when the rate or location of pair creation changes.  We propose to develop this model further, and to use it to discuss the visibility of the emission and the torque on the star, and how these change when the magnetosphere changes state.
 
\section*{Acknowledgments} 
We thank Dick Manchester for comments on the manuscript.

\appendix

\section{Response of a plasma at low frequencies}
\label{appendix:cold}

The response of a cold plasma may be described by the dielectric tensor $K_{ij}(\omega)$ \citep{Stix}:
\bea
&&D_i(\omega)=\varepsilon_0K_{ij}(\omega)E_j(\omega),
\nn
\\
\ms
&&K_{ij}(\omega)=
\left(\begin{array}{ccc}
S&-iD&0\\
iD&S&0\\
0&0&P
\end{array}
\right).
\label{pd3a}
\eea
At sufficiently low frequencies, when dissipation is neglected, one has
\be
S\approx1+{c^2\over v_A^2},
\qquad
D\approx0,
\qquad
P=1-{\omega_p^2\over\omega^2}.
\label{pd4}
\ee
The perpendicular component of the response is
\be
{\bi P}_\perp={c^2\over v_A^2}\varepsilon_0{\bi E}_\perp.
\label{pd6}
\ee
The time-derivative of (\ref{pd6}) gives
\be
{\bi J}_{\rm ind \perp}={c^2\over v_A^2}\varepsilon_0{\partial{\bi E}_\perp\over\partial t},
\label{pd7}
\ee
which reproduces (\ref{pd1}). The parallel response in the low-frequency, cold-plasma limit is included in (\ref{pd8_1}).

\section{Minimal charge density}
\label{appendix:min}

The charge density in the minimal model is given by (\ref{sc1}) with $E_{\rm ind\parallel}$ given by 
\be
E_{\rm ind\parallel}={\mu_0\over4\pi}
{{\bi b}\cdot[{\bi x}\times(\bomega_*\times{\bi m})]\over r^3},
\label{Amin1}
\ee
where we neglect the final (radiative) term in (\ref{Eind3}).  In terms of the angles introduced in (\ref{sc9}), one has
\be
E_{\rm ind\parallel}={\mu_0\omega_* m(\cos\theta-\cos\theta_m\cos\alpha)\over4\pi r^2\Theta},
\label{Amin2}
\ee
with $\Theta=(3\cos^2\theta_m+1)^{1/2}$. The unit vector along the magnetic field lines is
\be
{\bi b}=
{1\over\Theta}\bigg(2\cos\theta_m{\hat{\bi r}}
-{\partial\cos\theta_m\over\partial\theta}{\hat\btheta}
-{1\over\sin\theta}{\partial\cos\theta_m\over\partial\phi}{\hat\bphi}
\bigg),
\label{Amin3}
\ee
where ${\hat{\bi r}},{\hat\btheta},{\hat\bphi}$ are unit vectors for spherical polar coordinates relative to the rotation axis.
We note the identities
\bea
\sin\theta{\partial\cos\theta_m\over\partial\theta}=\cos\theta\cos\theta_m-\cos\alpha,
\qquad\qquad
\nn
\\
\ms
\left({\partial\cos\theta_m\over\partial\theta}\right)^2
+{1\over\sin^2\theta}\left({\partial\cos\theta_m\over\partial\phi}\right)^2=\sin^2\theta_m.
\label{Amin4}
\eea

An orthogonal coordinate system corresponding to dipolar magnetic field lines involves the coordinates \citep{Ketal05} $\mu,\chi,\phi_m$, with
\be
\mu=-{\cos\theta_m\over r^2},
\qquad
\chi={\sin^2\theta_m\over r},
\label{min16}
\ee
and $\phi_m$ the azimuthal angle relative to the dipole axis. The divergence of a vector, ${\bi V}$, in this coordinate system is
\bea
\div{\bi V}={1\over h_\mu h_\chi h_\phi}
\bigg[{\partial\over\partial\mu}(h_\chi h_\phi V_\mu)
\qquad\qquad\qquad
\nn
\\
+{\partial\over\partial\chi}(h_\mu h_\phi V_\chi)
+{\partial\over\partial\phi_m}(h_\mu h_\chi V_\phi)
\bigg],
\label{min18}
\eea
\be
h_\mu={r^3\over\Theta},
\qquad
h_\chi={r^2\over\sin\theta_m\Theta},
\qquad
h_\phi=r\sin\theta_m.
\label{min17}
\ee

We are concerned with a vector that only has a component along the magnetic field. One has
\be
\div({\bi b}E_{\rm ind\parallel})={\Theta^2\over r^6}
{\partial\over\partial\mu}\left[
{r^3\over\Theta}E_{\rm ind\parallel}
\right],
\label{min19}
\ee
where the partial derivative is at constant $\chi$, $\phi_m$. The derivative ${\bi b}\cdot{\rm grad}$ gives zero when operating on  $\chi$, $\phi_m$, and hence is proportional to this partial derivative. One has
\be
\left.{\partial\over\partial\mu}\right|_{\chi,\phi_m}
={{\bi b}\cdot{\rm grad}\over{\bi b}\cdot\grad\mu}
={r^3\over\Theta}\,{\bi b}\cdot{\rm grad},
\label{min20}
\ee
so that (\ref{min19}) reduces to
\be
\div({\bi b}E_{\rm ind\parallel})={\Theta\over r^3}
{\bi b}\cdot{\rm grad}\left[
{r^3\over\Theta}E_{\rm ind\parallel}
\right].
\label{min21}
\ee
The calculation is then lengthy but straightforward, leading to (\ref{sc10}).

\section{Screening current: RMM}
\label{appendix:Js}

The screening current,  ${\bi J}_{\rm cor}$,  in the RMM is associated with the time-derivative of $\rho_{\rm cor}$. It is in addition to the current $\rho_{\rm cor}\bomega_*\times{\bi x}$, and is given by
\be
{\bi J}_{\rm cor}=-\varepsilon_0\left[{\partial{\bi E}_{\rm cor}\over\partial t}+{\partial^2{\bi A}\over\partial t^2}\right].
\label{J1}
\ee
Explicit evaluation for a dipolar field gives
\be
{\bi J}_{\rm cor}=-{3\bomega_*\cdot{\bi x}\,\balpha\cdot{\bi x}\,{\bi x}
+r^2\balpha\cdot{\bi x}\,\bomega_*
+r^2\bomega_*\cdot{\bi x}\,\balpha
\over4\pi c^2r^5},
\label{J2}
\ee
with $\balpha={\bi m}\times(\bomega_*\times{\bi m})$. The component along the field lines is
\bea
{\bi J}_{\rm cor\parallel}=-{
\balpha\cdot{\bi x}(12\bomega_*\cdot{\bi x}\,{\bi m}\cdot{\bi x}
-r^2\bomega_*\cdot{\bi m})
\over4\pi c^2r^5[3({\bi m}\cdot{\bi x})^2+r^2m^2]}
\nn
\\
\times(3{\bi m}\cdot{\bi x}\,{\bi x}-r^2{\bi m}).
\label{J3}
\eea
The component across field lines, ${\bi J}_{\rm cor\perp}$, found by subtracting (\ref{J3}) from (\ref{J2}), has nonzero projections onto all of ${\bi x}$, ${\bi m}$, $\bomega_*$ and $\balpha$.

\end{document}